# A Comprehensive Survey on Nanophotonics Neural Networks

**Konstantinos Demertzis*[1,2], Georgios Papadopoulos[1], Lazaros Iliadis[2], Lykourgos Magafas[1]**

1   Department of Physics, Faculty of Sciences, International Hellenic University, Kavala Campus, St. Loukas, 65404, Greece; kdemertzis@tei-emt.gr; gebapad@teiemt.gr
2   School of Civil Engineering, Faculty of Mathematics Programming and General Courses, Democritus University of Thrace, Kimmeria, Xanthi, Greece; kdemertz@fmenr.duth.gr; liliadis@civil.duth.gr
*   Correspondence: kdemertz@fmenr.duth.gr; kdemertzis@teiemt.gr;

**Abstract:** In the last years, materializations of neuromorphic circuits based on nanophotonic arrangements have been proposed, which contain complete optical circuits, laser, photodetectors, photonic crystals, optical fibers, flat waveguides, and other passive optical elements of nanostructured materials, which eliminate the time of simultaneous processing of big groups of data, taking advantage of the quantum perspective and thus highly increasing the potentials of contemporary intelligent computational systems. This article is an effort to record and study the research that has been conducted concerning the methods of development and materialization of neuromorphic circuits of Neural Networks of nanophotonic arrangements. In particular, an investigative study of the methods of developing nanophotonic neuromorphic processors, their originality in neuronic architectural structure, their training methods and their optimization has been realized along with the study of special issues such as optical activation functions and cost functions.

## 1. Introduction

Artificial intelligence (AI) [1] enables machines to be trained so as to perform particular tasks, learn from experience, adapt to or interact with the environment and perform realistic anthropomorphic tasks [2]. Contemporary AI is one of the fastest evolving fields of information technology, in which high-level algorithmic approaches and tools descending from applied math's and engineering are used [3], [4] [5]. Most AI applications –from computers playing chess to automatically-driven cars– are based to a great extent on the intelligent technologies of Neural Networks [6] for the processing of multidimensional big data, with a view to revealing the hidden knowledge that is included in these groups [7], [8].

In classic von Neumann Architecture, where the computations are restrained by the speed of the channel between computation and memory (also known as the von Neumann congestion), even the important innovations on problems such as the shrinking of complete circuits, the reduction in their power needs and the decrease of temperature emitted by them cannot achieve the anticipated increases in their computing power [9], [10]. Even with the introduction of GPUs as extra processors for the improvement of graphic interface and the performance of tasks of high-level processing or the introduction of TPUs as the most powerful adapted processors for the performance of AI procedures, the capabilities of traditional systems seem unable to cope with the demands of modern technology and the uninterrupted flow of the data produced, even if they have offered some of the most important innovations [10]–[12].

The biggest challenge in this filed is the development of fully functional and utilizable Neuromorphic Systems-On-Chip (NSoC) [10], [13], which will be able to approach the biological human intelligence, performing the same tasks that the human brain effortlessly performs in no time at all and without remarkable consumption of resources and energy [14]. The neuromorphic computation comprises the creation of Neural Networks in matter, where the neurons of a physical device are connected with the corresponding synapses of physical devices [15]. The main motive for the neuromorphic computation is the time needed to process the computations and the energy performance provided by a distributed architecture which avoids the energy turbulence of data between the memory and the CPU [16].

The NSoCs based solidly on previous computational technology overcome the von Neumann congestion, massively use simultaneous computational procedures and are tolerant to faults [17]. Essentially, they form the way in which Neural Networks function conveying information in the same temporal and spatial way as the human brain. Furthermore, taking advantage of techniques such as the Memristors [10], they are capable of modelising learning skills; that is the adjustment ability of synapses in storing and conveying information depending on the evolution of a dynamic situation [16].

However, the most important development in the application efforts and standardization of NSoCs is spotted in the expanded efforts for developing fully Optical Neural Networks (ONN) [18], [19] and Photonic Neural Networks (PNN) [15], [20], [21]. The previously mentioned systems are based on the evolutions of optical technology and the most recent researches concerning photonics. Photonics is the science and technology field that deals with the creation, control and detection of photons, especially in the area of visible light and the near infrared electromagnetic spectrum (wavelength, polarization, transmission rate etc.) and the great potentials of their interconnection [20], [22], [23]. It is directly related, in basic as well as applied research, to quantum optics as to how linear transformations can be applied with the minimum energy consumption and with the slightest latency time on neuron level and with optoelectronics in the study of active and passive materials which interact electrically with light [24], [25].

## 2. Nature of Light

In general, the interaction of light with matter and its diffusion inside it is sufficiently described by optics (geometric and wave optics) [22], [26]. While studying various optic phenomena when the intensity of impacting radiations is small, the response in general and all the individual optic properties/parameters of materials (e.g. refraction index, absorption factor, polarization etc.) remain stable and independent of intensity [27]. However, when the intensity of radiation is high, as it happens with laser, for example, and in particular with focused-beam laser of great power, it has been proved experimentally that the optic response of matter and the optical parameters are modified, often significantly, and become dependent on the intensity of the radiation [28]. It is then that various extremely interesting phenomena take place, which are not detected in the case of low intensity impacting radiation [29]. These phenomena cannot be interpreted considering the linear response of matter, as it is expressed in the fundamental linear relation between the cause, i.e. the N electric field, and the result, i.e. the R inductive polarization [22], [30], [31]:

$$\vec{P} = \varepsilon_0 \chi^{(1)} \vec{E}$$

where $\chi^{(1)}$ is the linear susceptibility and $\varepsilon_0$ $\varepsilon$ is the intra-electrical invariant of vacuum. For the susceptibility $\chi^{(1)}$ and the refraction index n it is true that [24], [28]:

$$n^2 = \frac{\varepsilon}{\varepsilon_0} = 1 + \chi^{(1)}$$

In the case of high intensity radiation, conditions of high rate appear in the expression of polarization, the contributions of which are essential and cannot be omitted. These phenomena are the result of modification of the optic properties of matter due to the powerful electric field and the non-linearity of these phenomena is attributed to the fact that the response of matter is a non-linear function of the intensity of radiation [32].

One of the most important consequences of the linear response of matter under the influence of intense fields, apart from the alterations of the properties of matter, is that if there are different beams going through the same region of a non-linear medium simultaneously, they can interact with one another through matter [30]. Going this reasoning one step further and taking into consideration the principle of super-position, according to which a beam can be considered as the super-position of two beams of the same polarization, frequency and direction, we can assume that a beam can interact with itself [25], [26], [28].

As far as the procedure that causes the appearance of a non-linear optic behavior is concerned, when some radiation impacts on some material, it causes changes in the spatial and temporal distribution of the electric charge, inducing electric bipoles, the macroscopic result of which is the creation or modification of the polarization of the material [33]. For low values of the E field, the P polarization is analogous to the E field that caused it and the elementary bipoles, when oscillated, emit radiation of the same properties as the impacting radiation. Nevertheless, for high intensities of the E field, the radiation emitted by the elementary bipoles is not in correspondence with the electric E field that caused it. This can be explained by the fact that the captivated electrons of atoms/ions/unitary cells of crystal (or the structural unit in general) are forced to great displacements from their balance position. As a result, the motion of electrons cannot be described by the model of harmonic oscillator of Lorentz. Then, the radiation emitted contains frequencies different from those of the initial stimulating radiation. This practically means that it is possible to modify the impacting radiation itself with the addition of new frequencies, for example. In this way, non-linear phenomena can be interpreted and applied, which can explain how a beam of light can interact with one another (or with itself) creating amplification of light through light, merging of a beam with another one, production of new frequencies etc [34].

Based on what was previously mentioned, for high intensity of the electric field (e.g. for E>105 V/cm) where the presence of non-linear phenomena becomes significant, in the equation describing the polarization conditions of higher rate appear and the polarization is presented as an expansion of the Taylor sequence according to the following form [33], [35]:

$$\vec{P}(t) = \varepsilon_0 [\chi^{(1)} \vec{E}(t) + \chi^{(2)} \vec{E}^2(t) + \chi^{(3)} \vec{E}^3(t) + \cdots]$$

where $\chi^{(2)}$ is the second rate susceptibility, $\chi^{(3)}$ is the third rate susceptibility and so on and so forth. The susceptibilities are generally tensors, so, for instance, the first rate susceptibility $\chi^{(1)}$ is a second rate tensor with 3x3=9 elements and the corresponding polarization is presented by the following form [25], [34], [35]:

$$\begin{bmatrix} P_x^{(1)} \\ P_y^{(1)} \\ P_z^{(1)} \end{bmatrix} = \varepsilon_0 \begin{bmatrix} \chi_{xx}^{(1)} & \chi_{xy}^{(1)} & \chi_{xz}^{(1)} \\ \chi_{yx}^{(1)} & \chi_{yy}^{(1)} & \chi_{yz}^{(1)} \\ \chi_{zx}^{(1)} & \chi_{zy}^{(1)} & \chi_{zz}^{(1)} \end{bmatrix} \begin{bmatrix} E_x \\ E_y \\ E_z \end{bmatrix} \text{ or } P_i^{(1)} = \varepsilon_0 \sum_j \chi_{ij}^{(1)} E_j$$

With $i, j = x, y, z$.

Similarly, the non-linear second rate susceptibility $\chi^{(2)}$ is a third rate tensor $\chi_{ijk}^{(2)}$, whereas the third rate susceptibility $\chi^{(3)}$ is a forth rate tensor $\chi_{ijkl}^{(3)}$. In the case that the medium displays losses, the susceptibility $\chi^{(1)}$ is a complex quantity with its real part being connected to the linear refraction index n and its imaginary part being connected to the factor of linear absorption through the following relations [21], [22], [26]:

$$\chi^{(1)} = \text{Re}(\chi^{(1)}) + i[\text{Im}(\chi^{(1)})]$$

with $\text{Re}(\chi^{(1)}) \propto n_0$ and $\text{Im}(\chi^{(1)}) \propto \alpha_0$.

The equivalent relations also apply to the non-linear high-rate susceptibilities, which are also complex numbers with real and imaginary parts equivalent to the corresponding refraction indexes and absorption factors, which correspond to the non-linear refraction index and the non-linear absorption factor.

It is also important to point out that the calculation of an observed value in a system of photonic arrangements disrupts the system and, therefore, it shifts to a quantum condition in which the repetition of calculations of the same property leads to the same result. Thus, the following quantum conditions are observed [21], [24] [36]:

$|E\rangle$: Quantum condition where, if power is calculated, the result will be E.
$|p\rangle$: Quantum condition where, if momentum is calculated, the result will be p.
$|x\rangle$: Quantum condition where, if position is calculated, the result will be x.

In a general condition $|\psi\rangle$, the possibilities of calculating various physical properties are uncertain, that is there is the possibility $P1(E)$ of calculating the value of energy as E, $P2(p)$ is the possibility of calculating the value of momentum as p and so on and so forth. In a $|\psi\rangle$, condition system, after the calculation, for example, of energy with an E1, result, the wave function is disrupted and collapses (transforms) into a new condition $|E1\rangle$, so that the repetition of the same calculation gives the same result. Respectively, in a $|\psi\rangle$, condition system, after calculating for example the momentum with the result p1, the wave function is disrupted and collapses (transforms) into a new condition $|p1\rangle$, so that the repetition of the same calculation gives the same result [37] [38].

The conditions $|x\rangle$ and $|p\rangle$ cannot coincide, because the calculation of position (e.g. with photon scattering of short wavelength) alters the momentum. Consequently, there is no certainty about the momentum and the position of a particle as the values observed are random variables, in the sense that for every value of the spectrum of an observed physical quantity, a quantum width of probability for the calculation of this particular value corresponds to it. The total amount of quantum widths of probability for a spectrum of an observed physical quantity fully determines the quantum condition of the system. In that sense, one of the targets of photonic systems is the calculation of these widths of probability with the use of results of analytic methods [39], [40].

In conclusion, taking advantage of the binary nature of light and all the other characteristics which render it the fastest means of communication, the verge of modern investigation is focused on developing photonic neuromorphic processors [38], [41].

## 3. Photonic Neuromorphic Processors

The investigation into the development of photonic neuromorphic processors with passive optic circuits, focuses on advantages such as [21], [22], [24]:

Significant reduction of energy consumption in the applications of logical circuits as well as in data transfer.
Exceptionally high operating speeds with no energy consumption other than on the transmitters and the receivers.
Distribution of the computing power in the whole network, with each neuron performing simultaneously small parts of the whole computational activity.

Nevertheless, a big obstacle of photonic circuits has been the great volume of optical devices and the absence of susceptibility in contrast to the traditional integrated electronic circuits [13], [18], [21]. The materializations mainly on silicon (Si) and additionally on Indium Phosphide (InP) constitute a great innovative breakthrough in the materialization of integrated photonics, which is a reality nowadays. Especially, the materializations of integrated photonics with silicon as the construction material have proved excellent as they are transparent for wavelengths of 1270 to 1625 nm that are used in communications and the refraction index with a breadth of 3.48 in 1550 nm guarantees great resistance, while at the same time it can be checked thermally, electrically, mechanically or chemically [16], [17], [32].

Taking advantage of the properties above, silicon has been widely used for the materializations of passive elements, like waveguides, modulators, splitters, couplers and filters. On the other hand, Indium Phosphide allows the materialization of monolithically integrated solutions, which include a combination of passive and active devices, such as laser and amplifiers. Moreover, the integrated photonic technology offers the prospect of reducing the order of magnitude of the integration into nano-levels with all the significant advantages that the forementioned reduction of size brings about, such as the reduction of energy footprint, smaller size etc [40], [42].

Therefore, a multitude of arrangements of optical spare parts and integrated photonic circuits is already available, which results in the appearance of significant progress in the field of photonic neuromorphics, with the development of nanophotonic neuronic networks with either applications or waveguides or free-space optic [20], [36] [43].

## 4. Architectures

The philosophy behind the use of photonic circuits [22], [44], [45] in nano-arrangements [46]–[48] and materials is based on the need for significant improvement of the speed of transmitting and processing data and for an improvement of the energy efficiency of devices. The PNN materializations based on the fore mentioned materials, that have been present to this day, are classified into two main categories: with memory (Stateful) and without memory (Stateless), as they are concisely presented in figure 1 below [18], [21], [49]:

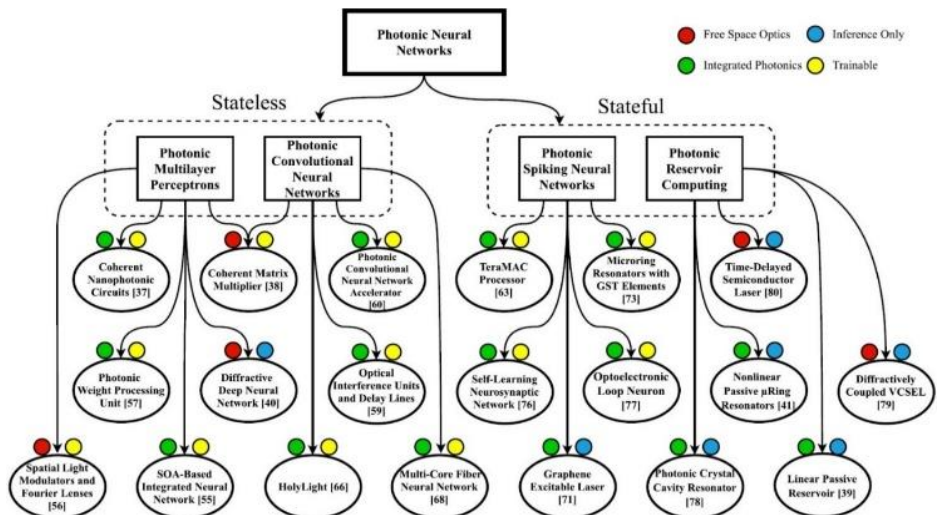

Figure 1: Stateful and Stateless Photonic Neural Networks

Moreover, the PNNs are classified according to design (waveguides or free-space optic) and according to optical training ability (Trainable) or reference only (Inference). The types of networks that have been applied in modern neuromorphic technology and their respective modifications are thoroughly explained below [18], [49].

*4.1 Perceptron*

The perceptron model consists of a single neuron, is the simplest autonomous system in existence and performs a particular task. This unique neuron of the system has a particular number of connections deriving from other neurons. The perceptron's development into ONNs is the most fundamental scientific field, with many articles having been published with respective materializations [20], [50], [51], [52], [53]. An All-Optical Neural Network (AONN) architecture with a hidden layer is presented in Figure 2 [52].

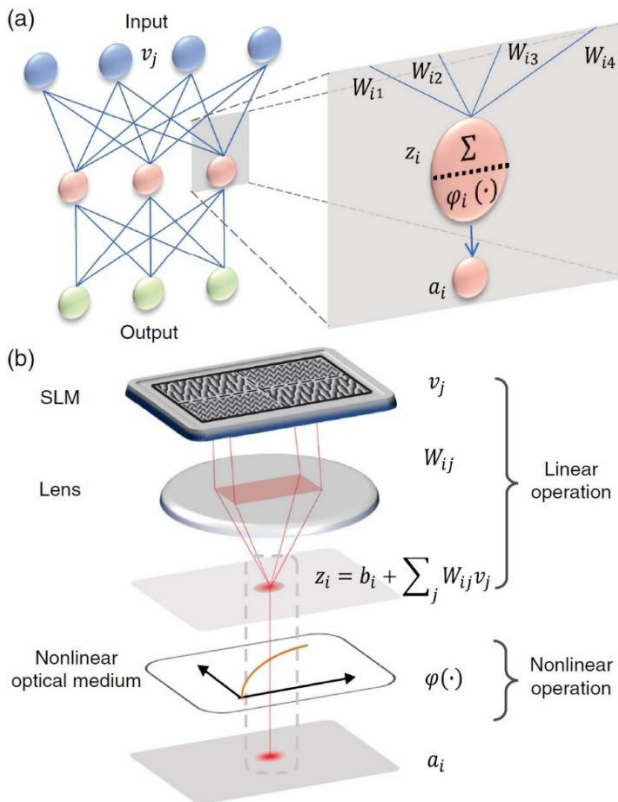

Figure 2: An ONN with full light use in all of its ranks.
It is based on free space optics, without the use of light waveguidance and integrated circuits, encoding the input signals with alterations in the illuminating power. During its linear operation, the light impacts on different areas of the surface of a Spatial Light Modulator (SLM) representing the knots $v_i$ of the input layer of a NN. With a special grid coating, the impacting light beam can be split in different j directions with weight Wij. The SLM is placed in the rear focal layer of the lenses, which apply Fourier Transformation and sum up all the diffracted beams on the focal point as follows [49], [51], [52]:

$$z_i = \sum_j W_{ij} v_j$$

as it happens with every knot of a conventional NN.
The non-linear operation is accomplished through Electromagnetically Induced Transparency (EIT) which is based on quantum phenomena and is produced by laser-cooled atoms 85Rb, in a Magneto-Optical Trap (MOT). The materialization of this particular architecture is shown in Figure 3 [40], [42], [54], [55]**Σφάλμα! Το αρχείο προέλευσης της αναφοράς δεν βρέθηκε.**:

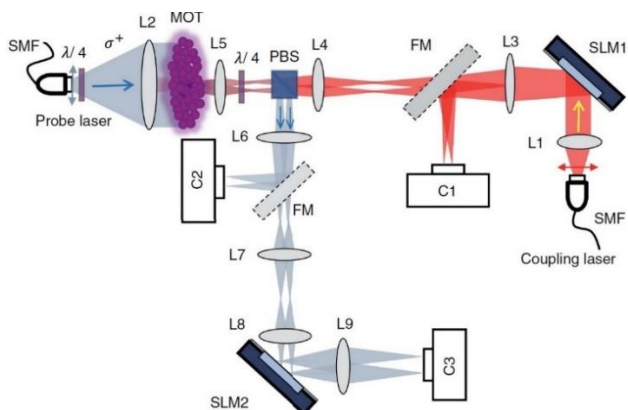

Figure 3: The optical arrangement in full development
The light beam from the Laser Single Mode Fiber (SMF), which constitutes the encoded input layer, is aligned (L1) and impacts on the first modulator (SLM1) which, in turn, emits four different beams. These are directed towards the L3 lens as the $v_j$, inputs, while at the same time the C1 camera, through a special mirror Flip Mirror (FM), records and calculates their values. Through the L4, L5 lens system the non-linearity is introduced by the MOT and afterwards the beams are directed towards the SLM2, after being recorded by the C2 camera first. Finally, the next layer (output layer), which consists of the SLM2 and the L7, L8, L9 lenses, transforms the four beams into two, which are recorded by the C3 camera [27], [29].

For the evaluation of this architecture, a classification of the different stages of an Ising model has been carried out, giving similar results compared to a NN created by a computer as these are represented in Figure 4 below [37].

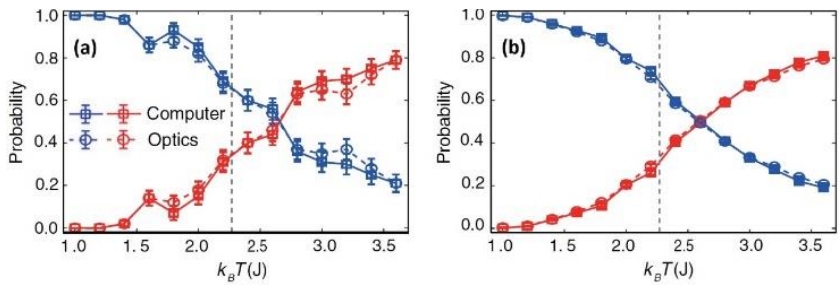

Figure 4: Average possibility of right (blue) and wrong (red) classification of this stage subject to temperature T (K) for 100 (on the left) and 4000 (on the right) settings.

It is obvious that this particular materialization can fully substitute a NN created by computers, with the only exception being the fact that commercial SLMs are not fast enough compared to a modern computer [39], [56].

*4.2 Multilayer Perceptrons*

A modified model of multiple sensors is called Multilayer Perceptron, in which between the input and output layers intervene one or more hidden layers. The data flow in such a network is always from left to right and there is no feedback loop. We also assume that the neurons in every layer interact only with those neurons that belong to their directly adjacent layers. In other words, the first hidden layer accepts the values of the input layer, the results of the first hidden layer go through the second hidden layer, whose results then go through to the third layer until they finally reach the final output layer. The materialization of the Nano-Photonic [46], [48] Multilayer Perceptron of Figure 5 is based on the use of nanophotonic circuits that process coherent light [49], [57]–[59].

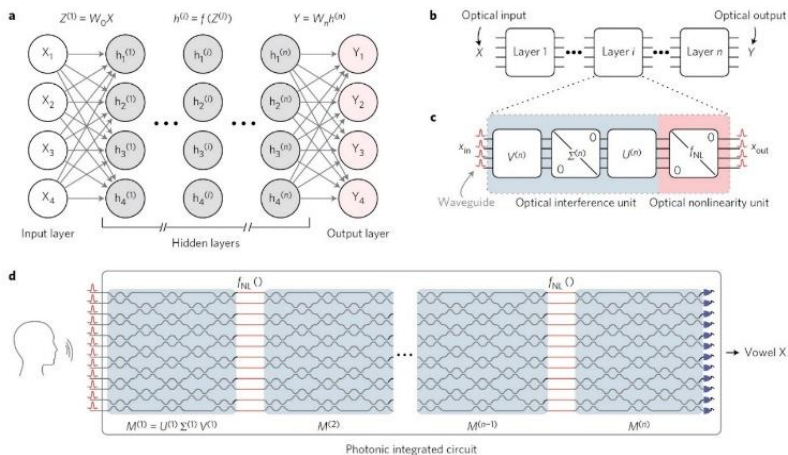

Figure 5. Nano-photonic Multilayer Perceptron

As shown in Figure 5, the basic theoretic block of NN (a) with the hidden layers (grey), is transferred to optical level operation (b), using two basic optical parts (c) and in particular an Optical Interference Unit (OIU) which performs multiplication of matrixes and an Optical Non-Linear Unit (ONU) which materializes the activation function. All the above are given briefly and in a integrated circuit form (d). The OIU consists of ranks of special programmed Mach-Zehnder interferometer (MZI). The MZIs convert the phase differences of light into amplitude differences (modulation). The structure of a MZI is shown in Figure 6 below [18], [51], [52], [57].

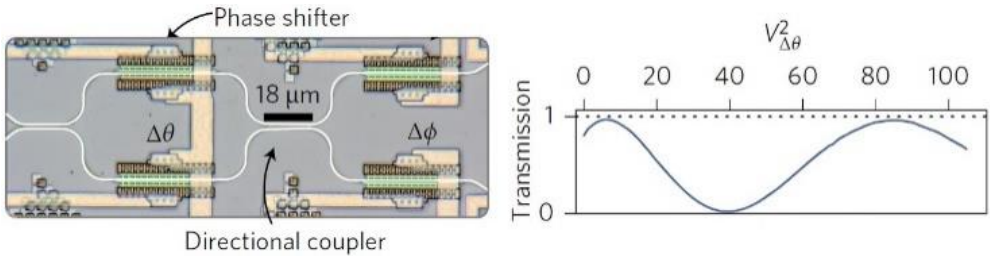

Figure 6: The programmed Phase shifter creates modifications in the phase which, in turn, are converted to amplitude modifications in the Directional coupler.

Finally, the modification of input M matrix of every i rank into a matrix product is accomplished as shown below [18], [49], [60]:

$$M^{(i)} = U^{(i)} \Sigma^{(i)} V^{*(i)}$$

according to the Singular-Value Decomposition (SVD), where $U$ $m \times m$ real or complex unitary matrix, $\Sigma$ $m \times n$ orthogonal diagonal matrix with no negative values in the diagonal and $V^*$ the conjugate transpose of V which is an $n \times n$ real or complex unitary matrix. It must be highlighted that through the M matrix, the weight matrix Wi of the NN is transferred to the optical circuit [22], [61], [62] [63].

The unit of the transfer function in this particular research paper has not yet been experimentally materialized and can only be simulated in a computer, with the transformation of signals from the optical to the electrical layer of operation and vice versa.

The experimental study and the training of ONN has been put into practice on a computer first, initially through the application of voice recognition with 76.7% accuracy. Then, the already familiar diagnostic tool of digital identification MNIST has been used, in which accuracy of results reach 95%, with the highest known value being 99%. This last conclusion shows the potentials and dynamics of Mechanical Learning in this particular field [18], [48], [52].

*4.3 Deep PNN*

The deep NNs are used in solving complex problems of high complexity. However, as the number of layers increases, its structure becomes more complex and this results in the input of a great computational load on the processor. Consequently, the training time increases and so does the energy consumption. These restrictions created the need for the materialization of ONNs of many layers DPNN, since the advantages of photonic transmission speed and the minimum energy consumption are indisputable [47], [49], [52] [64].

The construction of standardized, fully optical circuits, TNN of many layers is a true challenge nowadays. An arrangement for the materialization of a DPNN is shown in the following Figure 7 [18], [40], [49].

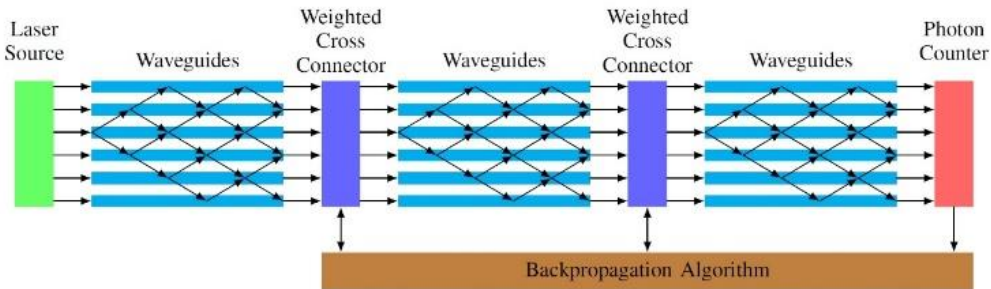

Figure 7. The architecture of a Deep Photonic Neural Network.

In this particular architecture the layers of NN are substituted with photonic grids, in which instead of knots with neurons we have waveguides. The interconnection between layers is accomplished through coupling devices with Weighted Cross Connectors so that the desired output from the network can be achieved. The coupling devices, which are responsible for the control of photons, consist of Optical Splitters and Optical Combiners for which the following relation is in effect [18], [22], [46], [52]:

$$c_i = \sum_j w_{ij} \cdot s_j$$

The weights $W_{ij}$ are controlled by external parameters until the network reaches the ideal output and be led to a condition of stable weights (training), where [19], [42], [49], [52]:

$$\alpha_i^{(l)} = \sum_{j=1}^{N_{(l-1)}} w_{ij} f_j^{(l-1)}(a^{(l-1)})$$

*4.4 Convolutional Neural Networks*

The Convolutional Neural Networks (CNN) [65] adopt a different approach in their organization as they take advantage of the hierarchical standard of the input data, creating more complex but fewer and simpler patterns in their architecture. The architecture of a CNN is analogous to the one of the connectibility pattern of neurons of the human brain and was inspired by the organization of the optical cortex. More analytically, a CNN is a deep learning algorithm which can take an image at the input, assign the appropriate weights to some of its various characteristics and, consequently, be able to differentiate one from the other. In other words, it has the ability to record successfully the spatial and temporary dependencies in an image through the application of relevant filters. Thus, a better adjustment to the total data is accomplished due to the decrease in the number of parameters that are involved and the reuse of weights. In other words, the network can be trained to better comprehend the structure of the image, while the preprocessing that is needed in a CNN is smaller when compared to other classification algorithms. The outcome is that CNNs have an advantage over the NN with perceptrons because the latter are prone to data overload due to the full connection of their knots.

There are several suggestions with CNNs that have been published such as [10], [53], [66]. A hybrid multilayer optical-electrical NN based on an optical matrix multiplier is presented in Figure 8.

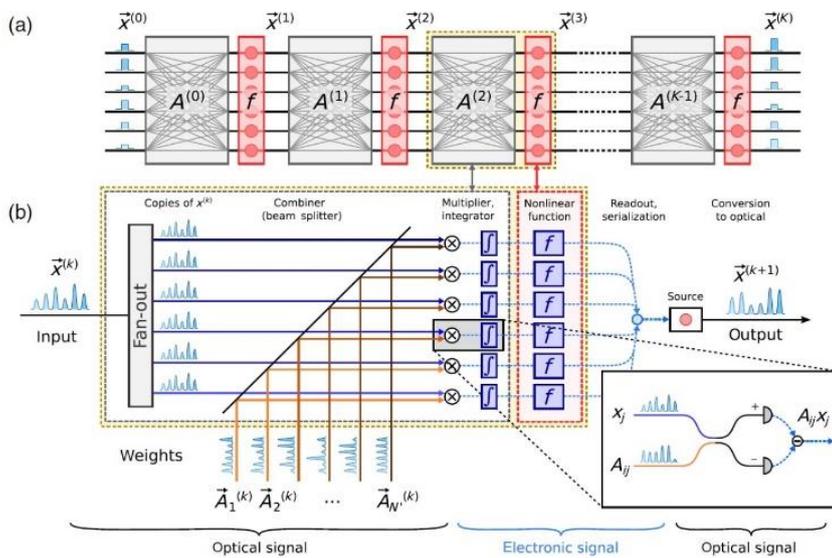

Figure 8: (a) Schematic diagram NN of K-layers consisting of a multiplier (grey) and an element for the activation function (red). (b) The multiplication performs a combination of inputs with the weight signals using homodyning.

In every one of the network layers the inputs $x^{(k)}$ are multiplied with the corresponding weights $A_i^{(k)}$, which are encoded as optical signals with homodyning between each pair of signal-weight. The electronic signals that derive are then subject to a non-linear transition function f and are converted to serial signals. Then, they are converted once again to optical signals and are sent to the input of the next layer. This optical system can be used for fully connected as well as CNNs and allows the inference of conclusions as well as the training in the same optical device.

Another suggestion of a CNN with full use of Optical Convolutional Neural Networks (OCNN) is presented in Figure 9 below [18], [49], [65], [67]:

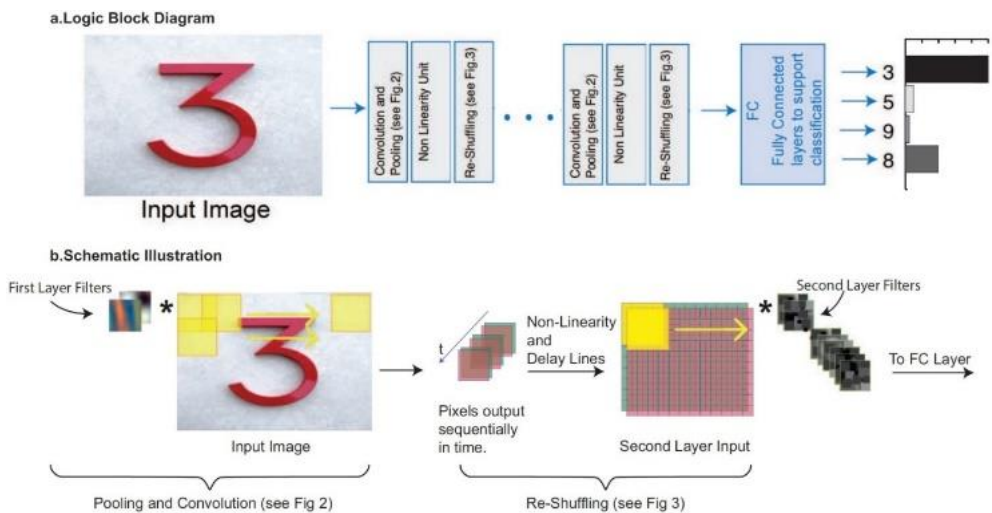

Figure 9: The suggested architecture for a fully-optical CNN.
The architecture consists of layers separated in an OIU based on MZI which performs Convolutional and Pooling, one part for the input of the Non-Linearity Unit and a splitters network of 3 dB, for the reorganization of data that the CNN is processing (Re-Shuffling).

The separators are programmed to introduce the appropriate time lag so that, at the output of the network layers, the signals could synchronize in time and form a new data entry for the input into the matrix nucleus of the next layer. It can be calculated that with this particular architecture the processing would be thirty times faster than that of an especially purpose-built electronic processor for CNNs with the same power consumption. As a result, such a system could play a significant role in the processing of thousands of terabytes of image and video data that are produced every day on the internet [41], [52].

*4.5 Spiking Neural Networks*

The Spiking Neural Networks (SNN) [68]–[70] are networks that imitate more than any other the biological NNs. Apart from the neural and synaptic condition, the SNNs incorporate the concept of time in their operating model. The idea behind this is that the neurons in a SNN should not trigger and be triggered in every propagation circle, as in standard networks of multiple layers with perceptrons. As it happens with the biological neurons, when the dynamics of their cell membrane reaches a particular value, which is called action potential, then the neuron triggers and produces a signal which travels to other neurons which, in turn, increase or decrease the dynamics of their cell membrane according to this particular signal. The SNNs use peak sequences as mechanisms of internal information presentation, in contrast to the usual continuous variables, while at the same time having equal, if not better, performance in computational cost to the traditional NNs [71]–[73].

In the field of photonic (Optical) SNNs many studies have been conducted in the past years [74], [75], initially taking advantage of the fast optical elements used in the construction of big systems with optical fibers. The successful applications finally led to the completion of arrangements, with the targets being greater expandability, increase of energy efficiency, reduction of cost and flexibility in the environmental fluctuations.

In a survey, the use of a graphene laser is recommended as an artificial neuron which is the fundamental element for the processing of information in the form of spikes. Moreover, the integrated layer of graphene is used as an optical absorber for the materialization of the non-linear activation function. The following Figure 10 presents the application with the use of circuits of free optics for the creation of a series of current peaks with adjustable characteristics of width and breadth [49], [76], [77].

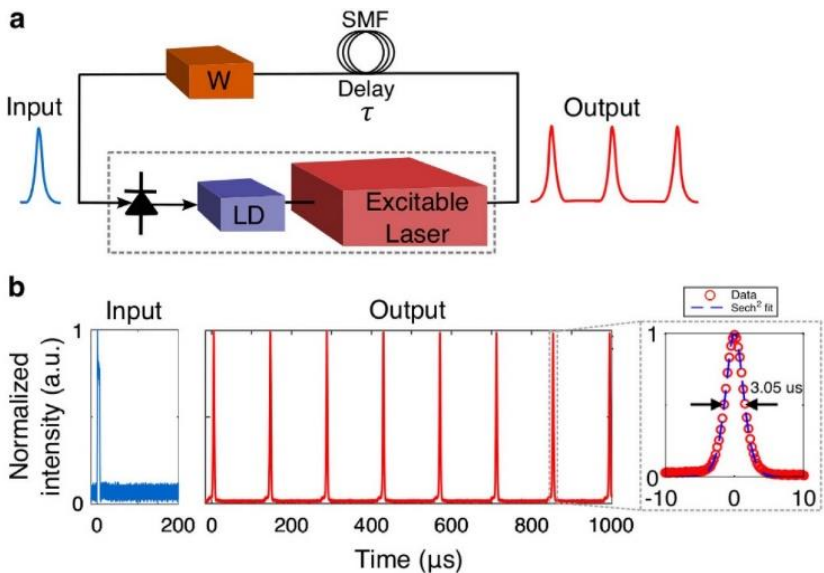

Figure 10: Circuit of creation of repeated current peak.

In another survey, the fundamental neuron is based on Distributed Feed Back (DFB) Laser of semi-conductors of Indium Phosphide. The use of this type of laser devices is very common in the construction of SNNs. The laser possesses two Photodetectors (PD), which allow inhibitory as well as excitatory stimuli. The recommended device is really fast reaching 1012 MACs/sec (MAC: Multiply Accumulate Operations) [78], [79].

*4.6 Reservoir Computing*

The use of Recurrent Neural Networks (RNN) [80], [81] has attracted researchers' interest because of their dynamics. The traditional RNNs, however, present some problems in training and designing, so an evolution has been suggested: the Reservoir Computing (RC). It is virtually, a neural network of feedback, where the input signals are dependent on time and present maximum efficiency compared to any other architecture in applications of sequence signals such as voice recognition, time series prediction etc. An RC system consists of a Reservoir through which a recording of inputs is conducted in a n-dimensional space and a Readout Layer, where the analysis of standards introduced in the reservoir is performed.

Optical applications with Photonics Reservoir Computing (PRC) Architecture have been presented in several research projects. In one of them passive optical elements are used for the materialization of the reservoir [50], [51], [82].

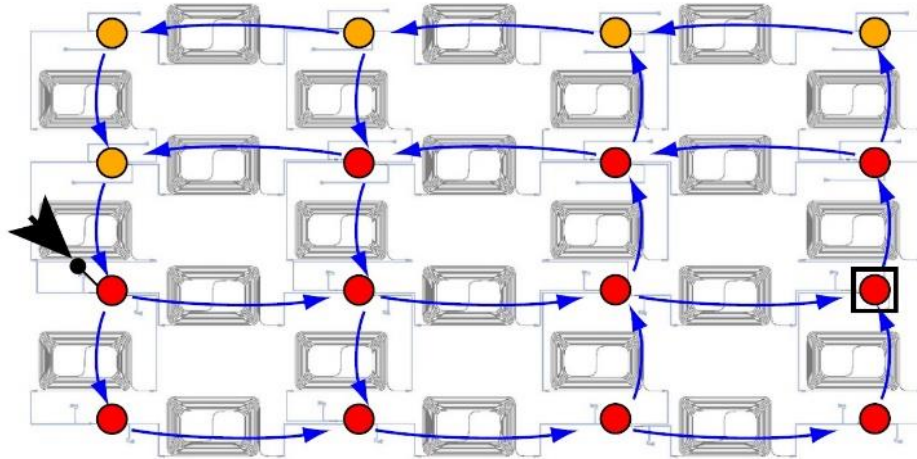

Figure 1: The reservoir structure in optical materialization. The black arrow shows the input.

The reservoir consists of a 4 x 4 = 16 knot system with splitters, couplers and waveguides, creating in this way a complex interferometer that operates in a random way. The fact that it comprises only passive elements renders it perfect from an energy efficiency point of view, but it displays solely a linear behaviour. This can be offset in the Readout layer with the introduction of a photodiode as a non-linear element [51], [82], [83].

Recently, a topology has been suggested for the Reservoir based on Micro rings Resonators (MR) which are non-linear elements and can cover the need for a non-linear transition function, simplifying in this way the Readout layer to the fullest extent.

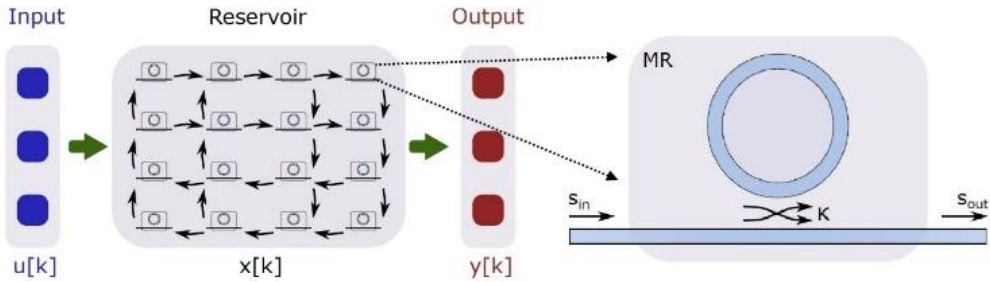

Figure 2: The reservoir with the 16 knots made from silicon on insulator (SOI) MR.

This particular topology has displayed a better error rate compared to others where the reservoir consists of passive linear elements.

An alternative suggestion for PRC is based on the use of photonic crystal cavity (PCC) is the shape of an ellipse quadrant. This particular architecture proves quite useful for projects of processing optical signals dependent on memory such as the header recognition of digital signals   [40], [67], [82].

## 5. Training Methodologies

Training is an important aspect of the Neural Networks since it does not only influence the behaviour of the network but also its overall efficiency. In supervised training, the training procedure uses an objectively calculated operation, where the distance (or error) between the desired and the real value is calculated. This operation is used to regulate the internal parameters of the NN that is the synapses' /connections' weights of neurons. In order to minimize the deflection between desired and real value, a gradient vector is calculated so that the way of how the error is influenced by any weight shift can be assessed [49], [52], [75].

Every time there is a change in the nature of data input in the network, the network needs to be retrained. This retraining can be done gradually as the network performs inference (online learning) [75] or it can be done independently, so that the network can adapt to a new input of training data (offline learning).

Given that the training includes gradient calculation, or even more complex calculations, it is a stage of resources and time consumption. In contrast, the inference (Inferences - that is the classification stage by the NN) is a much simpler procedure since the weights are already known in this stage. For this reason, many materializations PNN support only the inference stage and the weights are taken with the use of software applications on the level of electronic operation. Moreover, some applications cannot be trained at all, as in [40], [42], [84], [85]. These architectures are very fast and efficient as far as energy consumption is concerned, but they are not flexible, as they are especially designed for specific applications as their weights consolidate in the material during their construction.

When in a NN the training is electronic, two main disadvantages appear and, in particular, the physical system dependence on the accuracy of the model is added and the improvement of speed and the efficiency already accomplished with the optical part is eliminated. In order for the training, though complicated, to take full advantage of the photonic technology, it must be specifically adapted to optical architectures.

### 5.1 Propagation

The ONNs offer many advantages as far as the training of TNNs is concerned. In a conventional computer, the training is done using the Backpropagation Error Method and the Gradient Descent Application. Nevertheless, in some TNNs where the active number of parameters (parameters which are being calculated in every circle) surpasses by far the number of distinct parameters (as in RNNs and CNNs), the training with Backpropagation is definitely ineffective. In particular, the repeated nature of RNNs virtually makes them an extremely deep TNN, whereas in CNNs something relevant happens, since the same weight parameters are used repeatedly in different parts of an image for the output of its characteristics [18], [41], [86].

For the training of the network with Forward propagation, for the calculation of gradient in a particular modification step $\Delta_{w_{ij}}$ of the weight $w_{ij}$ of a NN, calculations of quantities are needed using Finite Difference Method (FDM) [87]–[89]:

$$f(w_{ij} + \delta_{ij})_{και} f(w_{ij})$$

και and after these two arithmetical operations, we calculate the weight change as follows:

$$\Delta w_{ij} = \frac{f(w_{ij} + \delta_{ij}) - f(w_{ij})}{\delta_{ij}}$$

In a conventional computer, the above procedure is computationally costy. On the other hand, in the field of photonic applications there are suggestions in ONNs which are better at the immediate calculation of the gradient, as every one of the fore steps of propagation is calculated in stable time, which is restricted only by the rate of photo detection which reaches 100GHz and the energy consumption is analogous only to the number of neurons.

This particular architecture is capable of reaching performance rates similar or even faster than Backpropagation with conventional computers (e.g. in very deep RNNs). Moreover, with the training procedure in the material (on chip) one can easily parameterize and train unitary matrixes, an approach which is particularly useful in deep Neural Networks [66].

Furthermore, in ONNs there is a possibility of training with the Backpropagation method based on an architecture where OIU with MZI are used for the linear operations of multiplications of matrixes. The algorithm of Backpropagation training generally operates in a circular mode between two stages, where in the first stage the error propagation is from the end of the network to its beginning and in the second stage there is a recalculation of the weights to check the contribution of each one to the output of the network.

In optical materializations, some basic restrictions to the control of weights are present, which have to be taken into account $w_{ij} \geq 0$. There cannot be a negative weight value since there is no negative light intensity value [42], [49], [66]:

$$\sum_i w_{ij}^{(l)} = 1$$

The initial light beam is split into waveguides so that the total of their intensities is stable. These particular restrictions are incorporated with the use of functions which transform the weights w to the desired breadth of activation function values, such as Softmax [52], [90], [91]:

$$w_{ij}^{(l)} = \frac{e^{w_{jj}^{(l)}}}{\sum_i e^{w_{ij}^{(l)}}}$$

In order for Backpropagation to be applied, the physical materialization of Adjoint Variable Method (AVM) [92], [93] is needed, which allows the reverse designing of photonic structures. According to this, at first, the adjoint of the initial field is created, the complementary one is propagated in the network reversely to the initial one and the initial field contributes with a replica of the reverse time of the complementary field. After all these, the conditions that yield gradient in every spot are expressed as the solutions of a classical conjugate electromagnetic problem and can be retrieved with an on-the-spot calculation of the field's intensity. A visualization of the operation of this particular method is presented in Figure 13 below [18], [41], [49], [86]:

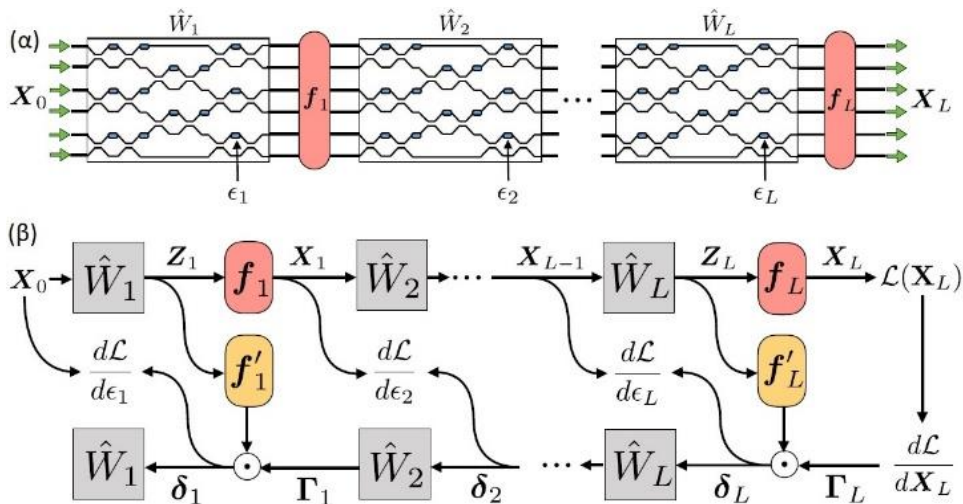

Figure 13: Backpropagation PNN

In stage (a) the squares correspond to the OIUs which materialize the linear operation (matrixes $W_L$). In blue colour we see the integrated phase shifters for the control of OIU and the training of the network. The red areas correspond to the non-linear activation functions $f_L$ which are performed through a computer. Respectively, in stage (b) the presentation of the operation for the calculation of TNN ranks. This route on top corresponds to the anterior propagation and the bottom to the Backpropagation.

This method allows the effective materialization of Backpropagation in a hybrid optical electronic network, with its main restriction being that a Forward feed system, which is mutual and with no losses, is necessary. Moreover, the fact that this method is based on classical Maxwell Electromagnetic Equations and not on a particular network form, renders it extremely flexible for its application on any photonic platform [49], [51], [65].

*5.2 Nonlinearity inversion*

In RC photonic applications the training concerns the Readout Layer [94], [95]. Recently, various researches on RC focused their attention on the development of the Reservoir with several recommended solutions. Nevertheless, the reading level is of fundamental importance because it determines ultimately the behaviour of the network and, unlike the Reservoir [51], [51], [82], must be appropriately trained. Hitherto, the training and the conjugation of signals on the reading level has been taking place in the conventional electric space and this resulted in the loss of any gain in speed and energy consumption that the optical part of arrangements introduced.

For a fully optical solution in the RC networks only a simple photo detector is required, which will receive the weighted total of all the optical signals. This approach, however, displays a drawback: we lose the ability for direct observation of the conditions of the photonic reservoir, which is necessary in many linear training algorithms. In order to solve this problem, there is a training procedure called Nonlinearity Inversion and is presented in the picture on the right [42], [96], [97].

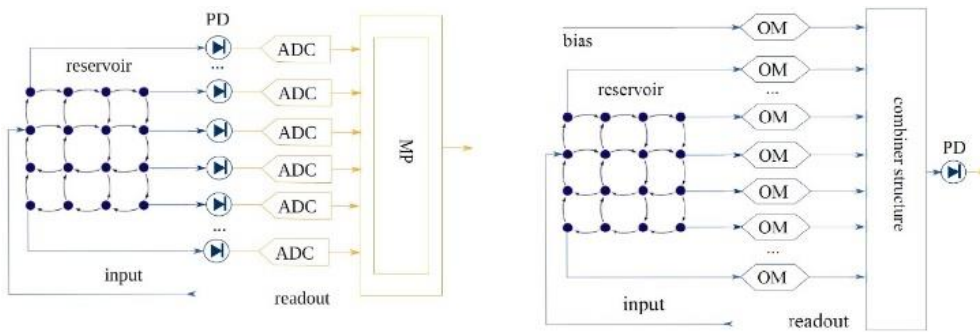

Figure 14: On the left: the signal from every knot is transferred through a photodetector (PD) to the electric space and through an AD converter to the microprocessor (MP). On the right: the optical signals (OM) are formed for the weight input, are added up (combine structure) and are finally transferred in the electric space.

This method solves the issue of direct observation of the Reservoir from the reading layer through a PD, with which a calculation of amplitude and its conditions' phase is materialized. The more complex conditions of the Reservoir are observed with the appropriate adaptation of the reading weights, whereas the feedback is achieved through a predetermined input sequence [47], [56], [80].

**6. Activation Functions**

Neurons are the structural element of the network. Each one of these knots receives a total of arithmetic inputs from different sources (either from other neurons or from the environment), performs some computation based on these inputs and produces an output. This output is either directed to the environment or constitutes input to the other neurons of the network. The computational neurons multiply each one of their inputs by the corresponding synaptic weight and calculate the total sum of the products. This total constitutes the activation function definition, which every knot materializes internally. The value that the function takes for this particular definition is also the output of the neuron for the current inputs and weights.

As a result, an important decision that has to be taken into account for the smooth operation of NNs is the selection of the activation function. In bibliographic references, the use of a powerfully non-linear function based on the electro-optical phenomenon is recommended for better results. Respectively, a plethora of non-linear functions have been materialized, which are presented below [80], [98].

*6.1 z–Transform (Complex Non-linearity)*

This function represents the $Z \rightarrow |Z|$ transformation and can be used for full, condense, polar mode. The bilateral z-Transform of a sequence of distinguishable time is defined as follows [99]–[101]:

$$X(z) = \sum_{n=-\infty}^{+\infty} x(n) \cdot z^{-n}$$

where the complex invariable z is called Complex Frequency and can be expressed with the use of polar coordinates. The z transformation of a sequence of distinguishable time is a total of infinite terms, which may converge to a real number for some values of the complex z variable and may not converge for some values of the complex z variable. The total of the variable values for which the z transformation exists, that is for which the total of z transformation converges, constitutes the Region of Convergence (ROC) [49], [102].

The reverse transformation is accomplished by calculating the reverse z transformations in each term of the total using z transformation pairs and, eventually, using the property of linearity of the z transformation. It is materialized with the method of analysis of the rational function in a total of simple fractions as is shown below [17], [49]:

$$X(z) = \frac{B(z)}{A(z)} = \frac{\sum_{k=0}^{M} b(k) \cdot z^{-k}}{\sum_{k=0}^{N} a(k) \cdot z^{-k}}$$

As an activation function in optical materializations, it is applied in signal analyses and, specifically, in solving linear equations of differences with fixed factors, in calculating the response and in designing linear filters or convolution layers [65].

*6.2 Electro Optic Activation (Complex Non-linearity)*

In NN applications of optical components, there is the possibility of creating non-linearity from the already existing material. The activation function is materialized converting a small part of the power of the input of the optic signal into electrical voltage. The remaining part of the initial optic signal is developed according to phase and amplitude by this voltage as it goes through an interferometer. A typical example of an electro-optical activation function is presented in Figure 15 below [56], [77], [82], [97]:

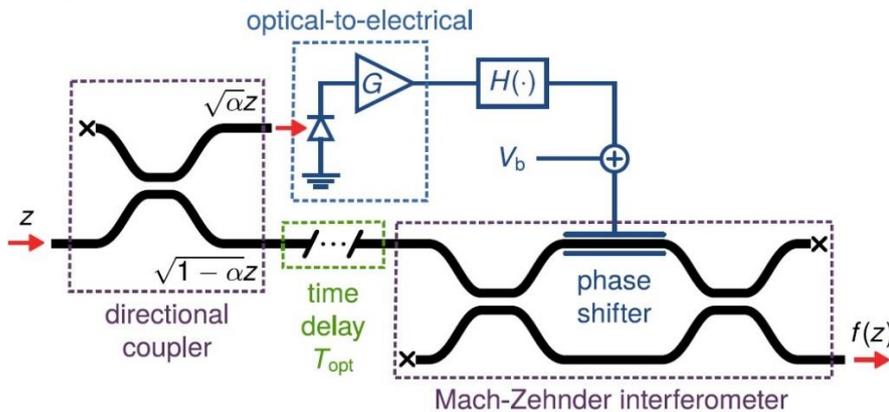

Figure 15: The arrangement for the electro-optical activation function.

For an input signal with a z value of amplitude, the non-linear activation function f(z) happens as the response of the interferometer and of the components throughout the route of the electric signal as is shown below [15], [41], [52]:

$$f(z) = j\sqrt{1-a} \cdot \exp\left(-j\left[\frac{g_\varphi |z|^2}{2} + \frac{\varphi_b}{2}\right]\right) \cdot \cos\left(\frac{g_\varphi |z|^2}{2} + \frac{\varphi_b}{2}\right) z$$

where: $\varphi_b = \pi \frac{V_b}{V_\pi}$ και $g_\varphi = \pi \frac{aGRR}{V_\pi}$

$\alpha$: the factor of input power transformation into an electric signal.
R: the response of the photodetector to the optical to electrical unit.
G: the gain of amplification rate.
Vb: the biasing voltage (bias)
Vπ: the required voltage for the π transformation of the phase.

*6.3 Sigmoid (Complex Non-linearity)*

The sigmoid activation function is used when a classification between two classes is needed or for a regression of weighted arrangements as it offers numbers between the space [-1, 1] at the output. This can be represented by the following transformation [56], [69], [75]:

$$z \to \frac{1}{1 + e^{-z}}$$

*6.4 Softmax (Complex Non-linearity)*

The Softmax function is represented by the transformation [90], [91]:

$$z \to \frac{e^z}{\sum e^z}$$

It is mainly used for multiclass problems.

*6.5 SPM Activation (Non-linearity)*

It represents the transformation [52], [103]:

$$Z \to Z \cdot e^{(-jG|z|^2)}$$

where $G = rad/\left(\frac{V^2}{m^2}\right)$ is the phase transformation for every unitary change of the input voltage.

*6.6 zReLU (Non-linearity)*

The zReLU is the Rectified Linear Units Function, with which the positive part of its definition is received as follows [10], [45], [49]:

$$f(z) = \begin{cases} z \text{ if Re }(z) > 0 \text{ and Im }(z) > 0 \\ 0 \end{cases}$$

*6.7 Cosine activation function (Non-linearity)*

Many of the recommended optical architectures for NNs use general-purpose equipment (e.g. for optical communications), whereas, ideally, they should be materialized inside a specific material (hardware). Consequently, there is no general approach to the training method of every recommended technique, as each one of them has its own characteristics that should be taken into account. A familiar problem that photonic architectures display concerns the activation function, due to the limited available choices and the difficulty of its materialization. Most of the suggestions use a combination of optical and electronic elements, such as the optical modulators Mach-Zender Modulator (MZM) [104], [105]. The result is a non-linear activation function of cosine form [56], [65], [68]:

$$P_{out} = P_{in}\sin^2\left(\frac{\pi}{2}\frac{V_{RF}}{V_\pi}\right)$$

where $P_{out}$ is the output signal, $P_{in}$ the Continuous Wave (CW) under modulation signal, $V_{RF}$ the input signal of the function and $V_\pi$ the value of input voltage for a phase shift of $\pi$ value.

Another important problem that has to be resolved in PNNs is the initialization of their parameters, such as the choice of the initial values of their weights. In their initial definition, the restrictions that exist in every materialization should be taken into consideration, as for example the constant bounded response of the signals that go through all the layers of the network. The topology with which an optical neuron of a cosine activation function is materialized, is shown in Figure 16 below [18], [49], [51], [52]:

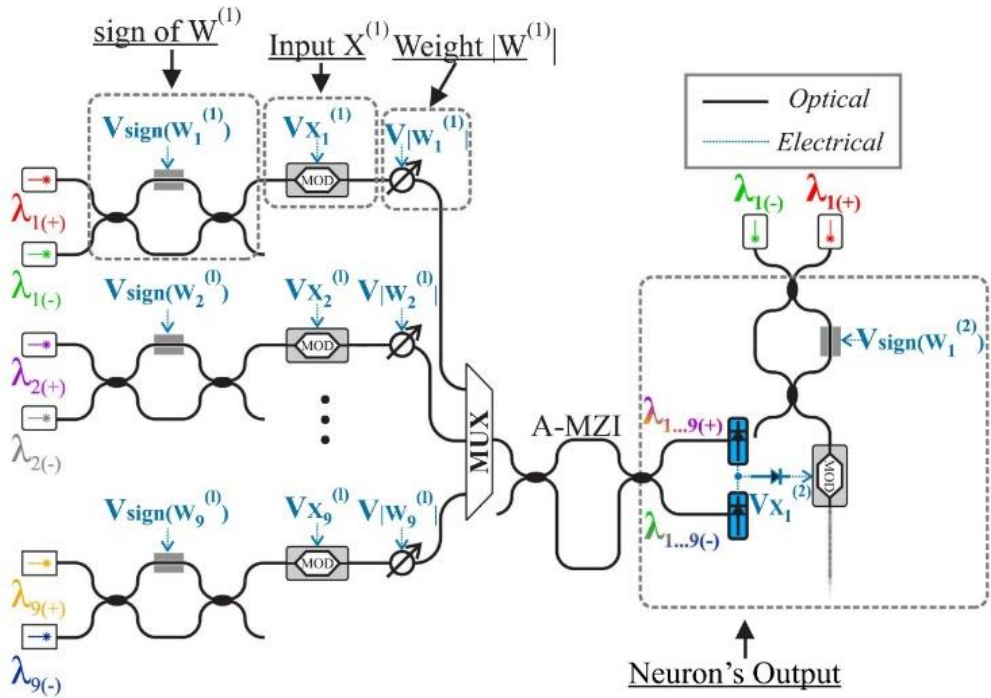

Figure 16: The operation principle of a neuron in an optical materialization.

In this particular materialization, two lasers of a different wavelength $\lambda i(+)$ and $\lambda i(-)$, are used, which, through the MZIs, function as switches (frame sign of W(1)) and correspond to positive and negative values of weights. Afterwards, the signals are led to the modulators (MOD-frame input X(1)) so that the input signal can be "printed" on an optical signal of power P(Xi(1)). The next level (frame Weight |W(1)|) includes a Variable Optical Attenuator (VOA) [106], [107] which is responsible for the amplification of signal-weight as is shown in the relation below [18], [40], [49]:

$$W_i^{(1)} \cdot P_{x_i^{(1)}}$$

In the next step, the signals are multiplexed (frame MUX) and are led in a grade of asynchronous MZI (A-MZI) for the separation of signals in signals of positive weight ($\lambda 1...9(+)$) and signals of negative weight ($\lambda 1..9(-)$) and in the end are added up in photodiodes (blue color). In conclusion, the MZM modulator that follows (MOD) and receives the two signals, operates in its non-linear area, materializing the transition function of cosine form. This particular architecture, where each neuron produces a signal which is led to the input of the next neuron, can be completed constructively and constitutes an independent photonic processor (chip) [20], [36], [76].

## 7. Conclutions

In this research paper, there has been an overview of the development and materialization methods of neuromorphic circuits of nano-photonic [57] arrangements for every respective contemporary architecture of conventional Neural Networks and the advantages and restrictions that arise during the transition from the electronic to the optical materializations were displayed. The fore mentioned networks are energy efficient, when compared to the corresponding electronic ones, and much faster due to photons. The reduction of simultaneous processing time increases radically the potentials of modern computational systems, which use optical arrangements, offering a promising alternative approach to micro-electronic and optical-electronic applications.

All these lead to the conclusion that there are potentials for a full transition to optical materializations as these display the following advantages:

Most of the systems do not require energy for the processing of optical signals. As soon as the Neural Network is trained, the computations on the optical signals are conducted without any additional energy consumption, rendering this particular architecture completely passive.

The optical systems, in contrast to the conventional electronic ones, do not produce heat during their operation and, as a result, they can be enclosed in three-dimensional constructions.

The processing speed in the optical systems is restricted only by the operation frequency of the laser source of light, which reaches 1THz.

The optical grids enable the multiplication of matrixes with vectors, something which is essential to NNs. The linear transformations (and some non-linear ones) can be performed at the speed of light and detected at a rate of over 100 GHz in photonic networks and, in some cases, with a minimum power consumption.

They are not particularly demanding as far as non-linearities are concerned, since many innate optical non-linearities can be used directly for the application of non-linear operations in PNNs, such as the activation functions.

On the downside, there are some difficulties in the transition to completely PNNs, which are the following:

The dimensions of optical devices are analogous to the light wavelength that they use (400nm – 800nm).

The mass production of optical devices is limited compared to the electronic ones, since they lack at least 50 years of research and development.

The training of the optical grids is quite difficult because the controlled parameters are active in matrix elements deriving from powerful non-linear functions.

The application of matrix transformations with optical components of mass production (such as fibers and lenses) is a restriction to the spread of ONNs due to the need for stability in the signal phase and to the huge number of neurons which are required in more complex applications.

Summarizing, although there are potentials concerning the materialization of PNNs, there are still some areas which require further research, such as some specific architectures of Deep Neural Nets (Long-Short-Term Memory Neural Networks, Generative Adversarial Nets, Geometric Deep Neural Networks, Deep Belief Networks, Deep Boltzmann Machines, etc.). Due to the significance of DNNs and the role they play in mechanical learning techniques, the research studies should focus on the question whether every type of conventional DNN can be converted in PNN, performing better and, thus, offering more advantages when compared to electronic arrangements. The ultimate goal in this is to replace the huge energy consuming NNs, with thousands of knots and multiple interconnections among hidden layers, with very fast optical arrangements.

There are also fields where the research on PNNs should focus on, such as the Hyper dimensional Learning (HL) [108], [109], a modern and very promising approach to NNs, which is still in the development stage. Here, the problem of a photonic materialization lies in the very big size of the internal representation of objects that are used in HL.

A further point that needs to be studied is the application of non-linear functions, which in most of the suggestions are materialized through software outside the optical arrangement. This results in the decline of performance, sometimes of a high rate, given that in multilayer NNs it is necessary to insert non-linearity many times successively.

Many more challenges need to be overcome, such as the many different hardware platforms that have been recommended, which are still under investigation with no clear winner yet. Moreover, we have to improve the already developed hardware as, in many cases, basic elements are still simulated or classic electronic ones are used. Furthermore, a critical element in a recommended NN architecture is its expandability in various applications, something which must be confirmed with further research studies. Finally, the field of NNs, which is still in early stage, is the massive integration of optical arrangements and, of course, their mass production, which is the last and most fundamental fortress of conventional NN arrangements against the transition to fully optical circuits.

Author Contributions: Conceptualization, K.D. and G.P.; methodology, K.D.; validation, K.D., G.P. and L.I.; formal analysis, G.P.; investigation, K.D. and G.P.; writing—original draft preparation, G.P.; writing—review and editing, K.D., G.P. and L.I.; visualization, G.P.; supervision, K.D.; project administration, L.I.; funding acquisition, K.D. All authors have read and agreed to the published version of the manuscript.
Funding: This research received no external funding.
Institutional Review Board Statement: Not applicable.
Informed Consent Statement: Not applicable.
Conflicts of Interest: The authors declare no conflict of interest.

The optical grids enable the multiplication of matrixes with vectors, something which is essential to NNs. The linear transformations (and some non-linear ones) can be performed at the speed of light and detected at a rate of over 100 GHz in photonic networks and, in some cases, with a minimum power consumption.

They are not particularly demanding as far as non-linearities are concerned, since many innate optical non-linearities can be used directly for the application of non-linear operations in PNNs, such as the activation functions.

On the downside, there are some difficulties in the transition to completely PNNs, which are the following:

The dimensions of optical devices are analogous to the light wavelength that they use (400nm – 800nm).

The mass production of optical devices is limited compared to the electronic ones, since they lack at least 50 years of research and development.

The training of the optical grids is quite difficult because the controlled parameters are active in matrix elements deriving from powerful non-linear functions.

The application of matrix transformations with optical components of mass production (such as fibers and lenses) is a restriction to the spread of ONNs due to the need for stability in the signal phase and to the huge number of neurons which are required in more complex applications.

Summarizing, although there are potentials concerning the materialization of PNNs, there are still some areas which require further research, such as some specific architectures of Deep Neural Nets (Long-Short-Term Memory Neural Networks, Generative Adversarial Nets, Geometric Deep Neural Networks, Deep Belief Networks, Deep Boltzmann Machines, etc.). Due to the significance of DNNs and the role they play in mechanical learning techniques, the research studies should focus on the question whether every type of conventional DNN can be converted in PNN, performing better and, thus, offering more advantages when compared to electronic arrangements. The ultimate goal in this is to replace the huge energy consuming NNs, with thousands of knots and multiple interconnections among hidden layers, with very fast optical arrangements.

There are also fields where the research on PNNs should focus on, such as the Hyper dimensional Learning (HL) [108], [109], a modern and very promising approach to NNs, which is still in the development stage. Here, the problem of a photonic materialization lies in the very big size of the internal representation of objects that are used in HL.

A further point that needs to be studied is the application of non-linear functions, which in most of the suggestions are materialized through software outside the optical arrangement. This results in the decline of performance, sometimes of a high rate, given that in multilayer NNs it is necessary to insert non-linearity many times successively.

Many more challenges need to be overcome, such as the many different hardware platforms that have been recommended, which are still under investigation with no clear winner yet. Moreover, we have to improve the already developed hardware as, in many cases, basic elements are still simulated or classic electronic ones are used. Furthermore, a critical element in a recommended NN architecture is its expandability in various applications, something which must be confirmed with further research studies. Finally, the field of NNs, which is still in early stage, is the massive integration of optical arrangements and, of course, their mass production, which is the last and most fundamental fortress of conventional NN arrangements against the transition to fully optical circuits.

Author Contributions: Conceptualization, K.D. and G.P.; methodology, K.D.; validation, K.D., G.P. and L.I.; formal analysis, G.P.; investigation, K.D. and G.P.; writing—original draft preparation, G.P.; writing—review and editing, K.D., G.P. and L.I.; visualization, G.P.; supervision, K.D.; project administration, L.I.; funding acquisition, K.D. All authors have read and agreed to the published version of the manuscript.
Funding: This research received no external funding.
Institutional Review Board Statement: Not applicable.
Informed Consent Statement: Not applicable.
Conflicts of Interest: The authors declare no conflict of interest.

23 of 24

24 of 24